\documentclass[a4paper]{scrartcl}
\usepackage[utf8]{inputenc}
\usepackage{amsmath,amssymb}
\usepackage{natbib}
\usepackage[UKenglish,ngerman]{babel} 
\usepackage{bm}
\usepackage{enumitem} 
\usepackage{framed} 
\usepackage{hyperref}
\usepackage{multirow}
\usepackage{eso-pic} 

\DeclareMathOperator*{\argmin}{arg\,min\,}

\DeclareMathOperator{\median}{median}
\DeclareMathOperator{\MAD}{MAD}
\DeclareMathOperator{\MSE}{MSE}

\renewcommand{\mathbf}[1]{\ensuremath{\bm{\mathrm{#1}}}}

\newtheorem{alg}{Algorithm}

\AddToShipoutPictureBG*{%
  \AtPageUpperLeft{%
    \raisebox{-\baselineskip}{%
      \makebox[5pt][l]{\quad \small This is an Author's Accepted Manuscript of an article to be published in \textit{Computational Statistics}}}
 }}%

\title{\Large The shooting S-estimator for robust regression}
\author{\normalsize Viktoria Öllerer  $^{a}$$^{\ast}$\thanks{$^\ast$Corresponding author. Email: viktoria.oellerer@kuleuven.be \vspace{6pt}}, Andreas Alfons $^{b}$ and Christophe Croux $^{a}$\\\normalsize\vspace{6pt} $^{a}${\em{Faculty of Economics and Business, KU Leuven, Belgium}};\\\normalsize $^{b}${\em{Erasmus School of Economics, Erasmus University Rotterdam, The Netherlands}}}
\date{\vspace{-0.5cm}}

\begin{document}
\selectlanguage{UKenglish}
\maketitle


\begin{abstract}
\noindent To perform multiple regression, the least squares estimator is commonly used. However, this estimator is not robust to outliers. Therefore, robust methods such as S-estimation have been proposed. These estimators flag any observation with a large residual as an outlier and downweight it in the further procedure. However, a large residual may be caused by an outlier in only one single predictor variable, and downweighting the complete observation results in a loss of information.
 Therefore, we propose the \textit{shooting S-estimator}, a regression estimator that is especially designed for situations where a large number of observations suffer from contamination in a small number of predictor variables. The shooting S-estimator combines the ideas of the coordinate descent algorithm with simple S-regression, which makes it robust against componentwise contamination,  at the cost of failing the regression equivariance property.\\
\smallskip

\end{abstract}


\section{Introduction}
In robust statistics it is generally assumed that the majority of observations is totally free of contamination. Any observation that deviates from the model is as a whole flagged as an outlier, even if only one component of the observation is contaminated. In case only a small number of predictor variables cause the deviation from the model, a lot of information is lost through downweighting the whole observation. Therefore, it seems more appropriate to not consider whole observations as outliers but only those components that really deviate from the model. This is especially useful if the majority of observations is contaminated in only a small number of variables.  Imagine, for example, a regression setting where in every observation one single predictor variable is contaminated. Here the usual robust methods break down, as there is not one single clean observation. But the majority of the cells of the design matrix is still clean and thus the majority of the data is still clean. In this setting, it is more suitable to use techniques developed for cellwise contamination (componentwise contamination) rather than those developed for rowwise contamination.

\cite{Alqallaf} extend the rowwise contamination model to also cover cellwise contamination. They define the influence function and the breakdown point in this setting and derive them for some multivariate location estimators, showing that these cannot cope with cellwise contamination. For principal component analysis, \cite{VanAelst} develop a method based on pairwise correlation that can deal with cellwise contamination. The same authors propose versions of the Stahel-Donoho estimator based on Huberized outlyingness \citep[see][]{VanAelst3} and cellwise weights \citep[see][]{VanAelst2}.

In this paper we derive a regression estimator, called the \textit{shooting S-estimator}, that can cope with cellwise contamination. It combines the ideas of the coordinate descent algorithm ('shooting algorithm') \citep[see][]{Friedman, Fu} with simple regression S-estimation \citep[see][]{Maronna}. In Section \ref{sec:objfunct}, we introduce the estimator. An algorithm is proposed in Section \ref{sec:algo}. We show simulation results in Section \ref{sec:sim} where we compare the shooting S-estimator to the least squares estimator and the robust S- and MM-estimators. Real data examples are presented in Section \ref{sec:realdata} and Section \ref{sec:conc} concludes.


\section{Motivation}
\label{sec:objfunct}

Our \textit{shooting S-estimator} uses the idea of the coordinate descent algorithm \citep[see][]{Friedman}, also called shooting algorithm \citep[][]{Fu}. Originally, this method performs, variable by variable, simple lasso regression. \cite{Tseng} showed that by iteratively looping through all variables, it converges to the lasso estimate for any starting value. However, it is well known that the lasso estimate is not robust \citep[see e.g.][]{Alfons}. In the shooting S-estimator, we achieve robustness by replacing the lasso estimation with unpenalized S-estimation \citep[see][]{Maronna}. In contrast to ordinary S-regression, the coordinate-wise approach of the coordinate descent algorithm allows us to weight all components of an observation differently.

The lasso estimate is defined as
\begin{align*}
\hat{\boldsymbol{\beta}}_{Lasso}= \argmin_{\boldsymbol{\beta}\in\mathbb{R}^p} \frac{1}{n}\sum_{i=1}^n (y_i-\sum_{j=1}^p x_{ij}\beta_j)^2+ 2\lambda\sum_{j=1}^p |\beta_j|.
\end{align*}
In the coordinate descent algorithm, to update the estimate of the lasso coefficient $\hat{\beta}_j$ $(j=1,\ldots,p)$, all other coefficients are kept fixed at $\hat{\beta}_k$ $(k\neq j)$
\begin{align}
\hat{\beta}_{j,Lasso}&= \argmin_{\beta_j\in\mathbb{R}} \frac{1}{n} \sum_{i=1}^n ((y_i-\sum_{k\neq j} x_{ik}\hat{\beta}_k) - x_{ij} \beta_j)^2+ 2\lambda\sum_{k\neq j} |\hat{\beta}_k|+2\lambda|\beta_j|\notag\\
&=\mathop{\mathrm{argmin}}_{\beta_j\in\mathbb{R}} \frac{1}{n}\sum_{i=1}^n ((y_i-\sum_{k\neq j} x_{ik}\hat{\beta}_k) - x_{ij} \beta_j)^2+2\lambda|\beta_j|.
\label{eq:lassoobj}
\end{align}
This can be seen as simple lasso regression where the new response
\begin{align}
{y}_i^{(j)} = y_i - \sum_{k\neq j}  x_{ik}\hat{\beta}_k,\quad i=1,\ldots,n,
\label{eq:yold}
\end{align}
is regressed on $x_{ij}$, for a fixed value of $j$.

\medskip

For the shooting S-estimator, we want to make sure that the new response $\tilde{y}_i^{(j)}$, to be defined below, is not influenced by outliers in the cells $x_{ik}$. Therefore, we first define regression weights
\begin{align}
w_{ik} = w(\frac{|\tilde{y}_i^{(k)} - x_{ik}\hat{\beta}_k|}{\hat{\sigma}_k})
\label{eq:cellweight}
\end{align}
where the argument of the weighting function $w(\cdot)$ is the residual of regressing $\tilde{y}^{(k)}_i$ on $x_{ik}$, scaled by a robust residual scale $\hat{\sigma}_k$. Thus, $w_{ik}$ determines the `outlyingness' of the cell $x_{ik}$ in the regression $\tilde{y}_i^{(k)}$ on $x_{ik}$.
The weighting function should be non-increasing on the positive numbers and take values in the interval  $[0,1]$.
Our preferred option - for reasons of simplicity -  is hard rejection, where $w(r)=1$ if $r\leq c$ and 0 otherwise. Choosing the cut-off value $c=3$, less than $0.3$\% of clean observations are expected to be flagged as outliers in the regression model with normal errors. Of course, other choices for the weight function are possible.

The new response is defined as
\begin{align}
\tilde{y}_i^{(j)}=y_i-\sum_{k\neq j} \tilde x_{ik} \hat{\beta}_k\ \ \ \text{ with } \tilde x_{ik}=w_{ik} x_{ik}+ (1-w_{ik}) \hat{x}_{ik}
\label{eq:ynew}
\end{align}
The difference with \eqref{eq:yold} is that in the computation of the new response the values $x_{ik}$ are replaced by a convex combination  $\tilde x_{ik}$ of the observed value $x_{ik}$ and of a `corrected' value $\hat{x}_{ik}$.
As we know $\tilde y_i^{(k)}$ and $\hat{\beta}_k$, this  `corrected' value $\hat{x}_{ik}$ is computed through calibration \citep{Brown}:
\begin{align}
\hat{x}_{ik}=\frac{\tilde{y}_i^{(k)}}{\hat{\beta}_k}.
\label{eq:xhat}
\end{align}
(To avoid computational problems, we set $\hat{x}_{ik}=0$ in case $|\hat{\beta}_k|$ is small.)
The $\tilde x_{ik}$ can be interpreted as a cleaned version of the cell  value $x_{ik}$ in the design matrix.
If an observation is flagged as an outlier and gets a zero weight,  the $\tilde x_{ik}$  equals the `corrected' value $\hat{x}_{ik}$. If an observation is declared as clean and gets a weight of one, the cleaned version equals the observed value.  Note that $\hat{x}_{ik}$ and $w_{ik}$ depend on $\hat{\beta}_k$, for $k \neq j$.

\medskip

To compute the regression estimate $\hat{\beta}_j$, we use instead of the lasso as in (\ref{eq:lassoobj}), the robust unpenalized simple S-regression estimator. This leads us to  the \textit{shooting S-estimator}, which is defined variablewise conditional on knowing the other estimates $\hat{\beta}_k$ with $k\neq j$,
\begin{equation}
\label{eq:objfunct}
 \hat{\beta}_j = \argmin_{\beta\in\mathbb{R}} \hat{\sigma}_j(\beta)\end{equation}
with  $\hat{\sigma}_j(\beta)$ defined as solution $s$ of the equation
\begin{equation} \label{eq:defMscale}
\frac{1}{n}\sum_{i=1}^n \rho(\frac{\tilde{y}_i^{(j)} - x_{ij}{\beta}}{s}) = \delta.
\end{equation}
Hence,  $\hat{\sigma}_j(\hat{\beta}_j)$ is an M-estimator of scale computed from the residuals.  Here $\delta$ equals the expected value of the $\rho$-function at the normal distribution, i.e. $\delta = \mathbb{E}[\rho(Z)]$ with $Z\sim\mathcal{N}(0,1)$. It is chosen such that the breakdown point of the estimator is not too low, while its efficiency is high enough. A higher value of $\delta$ implies a higher breakdown point, but a lower efficiency \citep[see e.g.][Chapter 3.4]{Rousseeuw}.

\smallskip

 As a $\rho$-function we will use either Tukey's biweight
\begin{align}
\rho_{BI}(z) = \begin{cases}
\frac{k_{BI}^2}{6}(1-(1-(\frac{z}{k_{BI}})^2)^3) &\text{ if } |z|\leq k_{BI}\\
\frac{k_{BI}^2}{6} &\text{ if } |z| > k_{BI},
\end{cases}
\label{eq:biweight}
\end{align}
or the skipped Huber
\begin{align}
\rho_{skH}(z) = \begin{cases}
\frac{1}{2} z^2 & \text{ if } |z| \leq k_{skH}\\
\frac{k_{skH}^2}{2} & \text{ if } |z| > k_{skH}.
\end{cases}
\label{eq:skH}
\end{align}
These two $\rho$-functions are quite different in nature. The skipped Huber loss is quadratic in a central region $[-k_{skH}, k_{skH}]$ and constant outside this interval. Thus, skipped Huber is a skipped version of the quadratic loss.  In contrast, the biweight loss is designed to be smooth while still bounding the effect of extreme values. Apart from those two loss functions any $\rho$-function \citep[see][p31, Def 2.1]{Maronna} could be used as well.

\medskip

The shooting S-estimator fulfills some natural equivariance properties. Assume a regression  model with intercept. The estimator is computed using the coordinate descent algorithm with starting value described in Section~\ref{sec:algo}.  If a constant $a$ is added to an explanatory variable, the corresponding estimate of the slope coefficient $\hat{\beta}_j$ stays unchanged, while the intercept shifts by $a\hat{\beta}_j$. If a constant $a$ is added to the response, none of the estimated coefficients changes and the intercept shifts by $a$. These properties can be shown using Equations~(\ref{eq:cellweight}), (\ref{eq:ynew}), (\ref{eq:xhat}) and (\ref{eq:objfunct}), as well as the properties of the proposed initial estimator. If a multiple $\gamma$ of an explanatory variable is added to the response, we would like the corresponding  slope coefficient to become $\gamma+\hat\beta_j$. This type of regression equivariance is fulfilled if the starting  value has this property. Since the proposed starting value uses Huberized values for the predictor variables, this property does not fully hold, although one could say that for the converged estimator it `practically' holds.


\section{Algorithm}
\label{sec:algo}

To compute the shooting S-estimate described in Section \ref{sec:objfunct}, we use an iterative procedure similar to the coordinate descent algorithm. We first describe the iteration steps, and afterwards the determination of initial values (see Algorithm~\ref{alg:Sest} for details). We assume that the model contains an intercept, denoted by $\alpha$.

\smallskip

{\it Loop.}  In each step of the coordinate descent loop (with fixed $j$), we calculate
the $\tilde{y}^{(j)}_i$ by \eqref{eq:ynew} and \eqref{eq:xhat} and then compute the simple regression S-estimate of
the $\tilde{y}^{(j)}_i$ on the $x_{ij}$. To do this, we use the iteratively reweighted least squares
(IRLS) algorithm recommended by \cite{Maronna}. It consists of
another iterative algorithm. In each iteration, called an I-step,  a weighted least squares estimate
of $\beta_j$ is calculated and subsequently, a new value of the M-estimator of scale $\hat{\sigma}_j(\hat{\beta_j})$ is computed by searching a fixed point of a recursive version of \eqref{eq:defMscale}, $f(s) = 1/(n\delta) \sum_{i=1}^n \rho((\tilde{y}_i^{(j)}-x_{ij}\hat{\beta}_j)/s)s=s$.

Although convergence of the coordinate descent loop is not assured, we have observed it empirically in all our simulations studies.

\smallskip

{\it Initial values.}
We  first Huberize the predictor values, and get  `approximately clean' predictors $\tilde x_{ij}^0$. Then we use  the  MM-estimator to get initial 	coefficients $\hat\beta_j^{(0)}$, with the  \textit{linear quadratic quadratic (lqq)} $\rho$-function \citep{Koller} and tuning constants set for 50\% breakdown point and 95\% efficiency.

\medskip

Algorithm \ref{alg:Sest} gives the details.  The code of the algorithm is available on the homepage of the first author.

\begin{table}
\begin{alg}
	\textbf{Computation of the shooting S-estimate for a regression model with constant term}
	\label{alg:Sest}
\end{alg}
\begin{framed}
	\footnotesize
	\begin{itemize}[nolistsep]
		\item[]\hspace{-0.5cm}\textit{\# Initialization}
		\item $L:= 0$ \textit{\# Number of steps in coordinate descent loop}
		\item $\tilde{x}_{ij}^{(0)}=\max(\median_i (x_{ij})-2\MAD_i(x_{ij}),\min(x_{ij},\median_i (x_{ij})+2\MAD_i(x_{ij})))$
		\item Compute {the slopes} $\mathbf{\hat{\beta}}^{(0)}$, {the intercept $\hat{\alpha}$ and the residual scale $\hat{s}$} from the MM-regression of $y_i$ on the Huberized predictors {$\tilde{x}_{ij}^{(0)}$} using the lqq $\rho$-function
		{ \item $\hat{\alpha}_j^{(0)}:=\hat{\alpha},$ $\quad j=1,\ldots,p$
			\item $s_j^{(0)}:=\hat{s},$ $\quad j=1,\dots,p$}
		{\setlength\itemindent{1cm}
			\item[]
			\item[]\hspace{-0.5cm}\textit{\# Coordinate descent loop}
			\item[$\diamond$] $L:= L+1$
			\item[$\diamond$] For $j=1,\ldots,p$ \textit{\# Index of the variable used in regression step}
			{\setlength\itemindent{2cm}
				\item[]
				\item[]\hspace{-0.5cm}\textit{\# Regression step}
				\item[$\centerdot$] $\tilde{y}_i^{(j)}:= y_i - \sum_{k< j}  \tilde{x}_{ik}^{(L)}\hat{\beta}_k^{(L)} - \sum_{k>j} \tilde{x}_{ik}^{(L-1)}\hat{\beta}_k^{(L-1)} ,$ $\hfill i=1,\ldots,n$
				\item[$\centerdot$] $r:= 0$ \textit{\# Number of I-steps}

				\item[$\centerdot$] $res_i^{(L,0)}:=\tilde{y}_i^{(j)} - x_{ij}\hat{\beta}_j^{(L-1)}-\median_i(\tilde{y}_i^{(j)} - x_{ij}\hat{\beta}_j^{(L-1)}),$ \\
				\item[] \hfill $\quad i=1,\ldots,n$
				\item[$\centerdot$] $\omega_{ij}^{(L,0)}:= {\rho'({res_i^{(L,0)}}/{s_j^{(L-1)}})}/({res_i^{(L,0)}}/{s_j^{(L-1)}}),$ $\hfill i=1,\ldots,n$
				{\setlength\itemindent{3cm}
					\item[]
					\item[]\hspace{-0.5cm}\textit{\# I-steps}
					\item[$\circ$] $r := r+1$
					\item[$\circ$] Compute {the slope} $\hat{\beta}_j^{(L,r)}$ {and the intercept $\hat{\alpha}_j^{(L,r)}$} from the
					\item[] weighted least squares regression of $\tilde{y}_i^{(j)}$ on $x_{ij}$ with
					\item[] weights {$\omega_{ij}^{(L, r-1)}$} \textit{\# $j$ is fixed}
					\item[$\circ$] $res_i^{(L,r)}:=\tilde{y}_i^{(j)} - x_{ij}\hat{\beta}_j^{(L,r)}-{\hat{\alpha}_j^{(L,r)}},$ $\quad i=1,\ldots,n$
					\item[$\circ$] $\ell:=0$ \textit{\# Number of M-steps to compute scale}
					\item[$\circ$] $s_0 = \begin{cases}
					\median_i|res_{i}^{(L,r)}|\cdot1.4826 &\text{ if } r = 1\\
					s_j^{(L,r-1)} &\text{ if } r>1
					\end{cases}$
					{\setlength\itemindent{4cm}
						\item[]
						\item[]\hspace{-0.5cm}\textit{\# M-step}
						\item[$\blacktriangle$] $\ell := \ell +1$
						\item[$\blacktriangle$] $s_\ell := \sqrt{\frac{s_{\ell -1}^2}{\delta\cdot n}\sum_{i=1}^n \rho(\frac{res_i^{(L,r)}}{s_{\ell-1}})}$
						\item[$\blacktriangle$]  Repeat M-step until $|\frac{s_\ell}{s_{\ell-1}} - 1|<\epsilon_1=10^{-6}$
					}
					\item[]
					\item[$\circ$] $s^{(L,r)}:=s_\ell$
					\item[$\circ$] { $\omega_{ij}^{(L,r)}:= \rho'(res_i^{(L,r)}/s^{(L,r)})/(res_i^{(L,r)}/s^{(L,r)}),$
						\item[] \hfill $\quad i=1,\ldots,n$}
					\item[$\circ$] Repeat I-step until $\max_i|res_i^{(L,r)} - res_i^{(L, r-1)}|<\epsilon_2$
					\item[] {\it \# $\epsilon_2=10^{-6} {\MAD_iy_i}$}
				}
				\item[]
				\item[$\centerdot$] $\hat{\beta}_j^{(L)}:=\hat{\beta}_j^{(L,r)}$
				\item[$\centerdot$] {$\hat{\alpha}_j^{(L)}:=\hat{\alpha}_j^{(L,r)}$}
				\item[$\centerdot$] $ s_j^{(L)}:=s^{(L,r)}$
				\item[$\centerdot$] $res_{i}^{(L)}:=res_{i}^{(L,r)}$ $\quad i=1,\ldots,n$
				\item[$\centerdot$] $\hat{x}_{ij}^{(L)}:=\left\{ \begin{array}{lll} (\tilde{y}_i^{(j)}-{\hat{\alpha}_j^{(L)}})/\hat{\beta}_j^{(L)} & \text{ if } |\hat{\beta}_j^{(L)}|\geq\epsilon_3 & \qquad i=1,\ldots,n \\
				\median_ix_{ij} & \text{ otherwise } & \end{array}\right.$
				\item[]  \# $\epsilon_3=10^{-4} (\MAD_iy_i)/(\MAD_ix_{ij})$
				\item[$\centerdot$] $w_{ij}^{(L)}:=w( res_i^{(L)}/s_j^{(L)} ) \qquad i=1,\ldots,n$
				\item[$\centerdot$] $\tilde{x}_{ij}^{(L)} := w_{ij}^{(L)}  x_{ij}+(1-w_{ij}^{(L)})\hat{x}_{ij}^{(L)},$ $\quad i=1,\ldots,n$
			}
			\item[]
			\item[$\diamond$] \textit{\# End for-loop}
			\item[$\diamond$] Repeat coordinate descent loop until $\sum_{j=1}^p| s_j^{(L)}-s_j^{(L-1)}|<\epsilon_4$
			\item[] {\it \# $\epsilon_4=10^{-2}\MAD_i y_i$}
		}
		\item $\hat{\beta}_j:=\hat{\beta}_j^{(L)}$
		\item $\hat{\alpha}:=\median_i (y_i - \sum_{j=1}^p  \tilde{x}_{ij}^{(L)}\hat{\beta}_j^{(L)})$
	\end{itemize}
\end{framed}
\end{table}


\section{Simulations}
\label{sec:sim}

To evaluate the shooting S-estimator, we compare it to the classical least squares estimator (LS), the ordinary S-estimator and the MM-estimator \citep[see][]{Maronna}. The shooting S-estimator is computed as in Algorithm~\ref{alg:Sest} once with the biweight $\rho$-function (\ref{eq:biweight}) and once with the skipped Huber $\rho$-function (\ref{eq:skH}). We choose $k_{BI} = 3.420$ and $k_{skH} = 2.177$. This corresponds to a breakdown point of $20$\% in the simple regressions.  Our choice seems to be a good trade-off between robustness and efficiency. In practice, the breakdown point needs to be increased if the data at hand is more severely contaminated than in this simulation setting. For the computation of the ordinary S-estimate, we use the biweight loss function and set again $k_{BI} = 3.420$. The MM-estimator is computed with the standard settings of $50$\% breakdown point and an efficiency of $95$\% at the normal model, using the biweight loss function. We stick here to the high breakdown point of $50$\%, as MM can achieve high efficiency and a high breakdown point simultaneously. Thus, lowering the breakdown point would not increase the efficiency of the MM-estimator.

For the simulation setup we take $n=100$ and $p=15$. The regression coefficients $\mathbf{\beta}$ are taken equally spaced over the interval [0,1],  i.e. $\beta_j = j/p$ for $j=1,\ldots,p$. The predictors $\mathbf{x}_i$ and errors $e_i$ are independent and identically normally distributed with mean 0 for $i=1,\ldots,n$. We choose two different sampling schemes, one with uncorrelated and one with correlated predictors. For the first one, we use the identity matrix as a covariance matrix for the predictors. The error variance is $\sigma^2 = 0.5^2$. In the correlated setting we choose the predictor covariance matrix $\Sigma$ with $\Sigma_{ij}=0.5^{|i-j|}$ and the error variance $\sigma^2 = 0.81^2$. By this the signal-to-noise ratio\footnote{The signal-to-noise ratio equals $\frac{\sqrt{\mathbf{\beta}'\Sigma\mathbf{\beta}}}{\sigma}$.} is the same in both settings. The response variable is then created as $y_i = \mathbf{x}_i'\mathbf{\beta} + e_i$ for $i=1,\ldots,n$.

To every generated data set, we add $1$\%, $2$\%, $5$\% and $10$\% of cellwise contamination. The cells $x_{ij}$ that we contaminate are chosen randomly from the design matrix $X$. So every cell of our data set is equally likely to be contaminated. Three different contamination settings are used: a dense cluster $x^{cont}_{ij}\sim\mathcal{N}(50, 1)$, scattered outliers $x^{cont}_{ij}\sim\mathcal{N}(0, 100^2)$ and a wide cluster $x^{cont}_{ij}\sim\mathcal{N}(50, 10^2)$. We only contaminate the $x$-values and not the $y$-values, which creates bad leverage points. For comparison, we also construct classical contamination settings where we choose whole rows for contamination instead of cells. For these we choose the three contaminations $\mathbf{x}_i^{cont}\sim\mathcal{N}(\mathbf{50},\Sigma)$, $\mathbf{x}_i^{cont}\sim\mathcal{N}(\mathbf{0},100^2\cdot\Sigma)$ and $\mathbf{x}_i^{cont}\sim\mathcal{N}(\mathbf{50},10^2\cdot\Sigma)$. Additionally, we also want to demonstrate that the shooting S-algorithm can deal with contamination in the response. From the clean data set, we select $1$\%, $2$\%, $5$\% and $10$\% of observations and generate their error terms as $e_{cont}\sim\mathcal{N}(50,\sigma^2)$ to create vertical outliers.

To compare the different estimators, we apply them to $R=1000$ generated data sets. For each data set, we compute the mean squared error (MSE)
\begin{align*}
\MSE(\mathbf{\hat{\beta}}) &= \frac{1}{p} \sum_{j=1}^p \frac{1}{R} \sum_{r=1}^R (\hat{\beta}_j^{(r)}-\beta_j)^2.
\end{align*}
Additionally, also the bias or the median squared error could be used as evaluation methods. We omit them as they are in line with the MSE.

The simulation results for cellwise contamination are displayed in Tables \ref{tab:res_cell} and \ref{tab:res_cell_MSE_corrpred} for uncorrelated and correlated predictors, respectively. Table \ref{tab:res_row_MSE_corrpred} gives the results for rowwise contamination in the data set with correlated predictors. Table \ref{tab:res_vertical} illustrates the behavior of the estimators in presence of vertical outliers for correlated predictors. The standard errors around the reported results are smaller than 4\% of the reported numbers in all tables. We omit the results for rowwise contamination and vertical outliers for uncorrelated predictors as they are comparable to the ones in the correlated case.
\medskip

\begin{table}[t]
\centering
\caption{\small{$n\cdot MSE$ of different estimators for cellwise contamination for all three contamination settings  with $n=100$, $p=15$ and uncorrelated predictors}}
\label{tab:res_cell}
{\footnotesize
\begin{tabular}{|c|r|rrrrr|}
  \hline
 & $\quad$ & $\epsilon = 0$ & $\epsilon = 0.01$ & $\epsilon = 0.02$ & $\epsilon = 0.05$ & $\epsilon = 0.1$ \\
  \hline
 & LS & 0.30 & 23.96 & 31.86 & 35.97 & 36.46 \\
   & S & 0.36 & 1.08 & 12.55 & 32.44 & 36.36 \\
   & MM & 0.33 & 0.48 & 0.88 & 18.35 & 34.52 \\
  $x_{ij}^{cont}\sim\mathcal{N}(50,1)$ &  &  &  &  &  &  \\
   & shooting S + BI & 0.43 & 0.62 & 0.84 & 1.72 & 5.37 \\
   & shooting S + skH & 0.55 & 0.65 & 0.80 & 2.02 & 5.61 \\
   \hline
 & LS & 0.30 & 22.41 & 30.82 & 36.16 & 36.67 \\
   & S & 0.36 & 0.99 & 11.15 & 31.21 & 36.54 \\
   & MM & 0.33 & 0.50 & 0.99 & 16.76 & 33.63 \\
  $x_{ij}^{cont}\sim\mathcal{N}(0,100^2)$ &  &  &  &  &  &  \\
   & shooting S + BI & 0.43 & 0.62 & 0.86 & 2.00 & 8.94 \\
   & shooting S + skH & 0.55 & 0.66 & 0.81 & 2.21 & 7.48 \\
   \hline
 & LS & 0.30 & 23.81 & 31.74 & 35.97 & 36.47 \\
   & S & 0.36 & 1.08 & 12.69 & 32.49 & 36.37 \\
   & MM & 0.33 & 0.49 & 0.92 & 18.50 & 34.41 \\
  $x_{ij}^{cont}\sim\mathcal{N}(50,10^2)$ &  &  &  &  &  &  \\
   & shooting S + BI & 0.43 & 0.62 & 0.84 & 1.76 & 5.72 \\
   & shooting S + skH & 0.55 & 0.65 & 0.81 & 2.05 & 5.83 \\
   \hline
\end{tabular}
}
\end{table}

For uncorrelated predictors, Table \ref{tab:res_cell} demonstrates the need of a new method that can deal with cellwise contamination. As well known, LS breaks down for any amount of contamination. But also the robust MM- and S-estimator have problems with larger amounts of cellwise contamination. As $2$\% of cellwise contamination corresponds in this setting to about $20-30$\% of rowwise contamination\footnote{The expected value of the number of contaminated rows is $n(1-(1-\epsilon)^p)$ for a cellwise contamination level $\epsilon$.}, the ordinary S-estimator already breaks down. As we have chosen a breakdown point of $50$\% for MM, it can deal with slightly higher contamination. But for about $5$\% of cellwise contamination it also breaks down. In contrast, the shooting S-estimators can deal with much higher levels of cellwise contamination. They are reliable even for up to 10\% of cellwise contamination, or around 80\% of rowwise contamination. The two shooting S-estimators perform comparably in this setting.

\begin{table}[t]
\centering
\caption{\small{$n\cdot MSE$ of different estimators for cellwise contamination for all three contamination settings  with $n=100$, $p=15$ and predictors with correlation matrix $\Sigma$}}
\label{tab:res_cell_MSE_corrpred}
{\footnotesize
\begin{tabular}{|c|r|rrrrr|}
  \hline
 & $\quad$ & $\epsilon = 0$ & $\epsilon = 0.01$ & $\epsilon = 0.02$ & $\epsilon = 0.05$ & $\epsilon = 0.1$ \\
  \hline
 & LS & 1.28 & 35.28 & 39.47 & 35.46 & 36.02 \\
   & S & 1.53 & 6.10 & 21.57 & 45.55 & 40.45 \\
   & MM & 1.39 & 2.82 & 5.58 & 26.88 & 46.73 \\
  $x_{ij}^{cont}\sim\mathcal{N}(50,1)$ &  &  &  &  &  &  \\
   & shooting S +  BI & 1.70 & 2.28 & 2.84 & 3.55 & 6.20 \\
   & shooting S + skH & 2.00 & 2.26 & 2.44 & 3.66 & 6.07 \\
   \hline
 & LS & 1.28 & 32.66 & 38.84 & 36.16 & 36.54 \\
   & S & 1.53 & 5.60 & 19.17 & 43.82 & 40.40 \\
   & MM & 1.39 & 2.93 & 5.66 & 25.74 & 44.23 \\
  $x_{ij}^{cont}\sim\mathcal{N}(0,100^2)$ &  &  &  &  &  &  \\
   & shooting S + BI & 1.70 & 2.32 & 2.95 & 4.25 & 10.03 \\
   & shooting S + skH & 2.00 & 2.29 & 2.53 & 4.04 & 7.74 \\
   \hline
 & LS & 1.28 & 34.98 & 39.24 & 35.47 & 36.04 \\
   & S & 1.53 & 6.15 & 21.62 & 45.52 & 40.38 \\
   & MM & 1.39 & 2.90 & 5.55 & 27.17 & 46.51 \\
  $x_{ij}^{cont}\sim\mathcal{N}(50,10^2)$ &  &  &  &  &  &  \\
   & shooting S + BI & 1.70 & 2.28 & 2.86 & 3.63 & 6.67 \\
   & shooting S + skH & 2.00 & 2.25 & 2.45 & 3.69 & 6.23 \\
   \hline
\end{tabular}
}
\end{table}

Table \ref{tab:res_cell_MSE_corrpred} confirms for correlated predictors what is shown in Table \ref{tab:res_cell} for uncorrelated ones. The only major difference is that for correlated predictors the shooting S-estimators already outperform the MM-estimator for 1\% of cellwise contamination, even though the MM-estimator does not break down yet in this case.

\begin{table}[t]
\centering
\caption{\small{$n\cdot MSE$ of different estimators for rowwise contamination for all three contamination settings  with $n=100$, $p=15$ and predictors with correlation matrix $\Sigma$}}
\label{tab:res_row_MSE_corrpred}
{\footnotesize
\begin{tabular}{|c|r|rrrrr|}
  \hline
 & $\quad$ & $\epsilon = 0$ & $\epsilon = 0.01$ & $\epsilon = 0.02$ & $\epsilon = 0.05$ & $\epsilon = 0.1$ \\
  \hline
 & LS & 1.28 & 53.62 & 53.96 & 54.72 & 54.80 \\
   & S & 1.53 & 1.51 & 1.50 & 1.48 & 1.48 \\
  $\mathbf{x}_{i}^{cont}\sim\mathcal{N}(\mathbf{50},\Sigma)$ & MM & 1.39 & 1.40 & 1.41 & 1.44 & 1.50 \\
   &  &  &  &  &  &  \\
   & shooting S + BI & 1.70 & 1.66 & 1.63 & 1.63 & 1.66 \\
   & shooting S + skH & 2.00 & 1.91 & 1.88 & 1.77 & 1.78 \\
   \hline
 & LS & 1.28 & 12.58 & 22.82 & 44.10 & 56.41 \\
    & S & 1.53 & 1.57 & 1.58 & 1.75 & 2.27 \\
  $\mathbf{x}_{i}^{cont}\sim\mathcal{N}(\mathbf{0},100^2\Sigma)$ & MM & 1.39 & 1.43 & 1.47 & 1.58 & 1.79 \\
   &  &  &  &  &  &  \\
   & shooting S + BI & 1.70 & 1.69 & 1.73 & 1.98 & 3.04 \\
   & shooting S + skH & 2.00 & 1.95 & 1.94 & 1.98 & 2.48 \\
   \hline
 & LS & 1.28 & 56.43 & 56.28 & 55.85 & 49.89 \\
   & S & 1.53 & 1.51 & 1.50 & 1.48 & 1.48 \\
  $\mathbf{x}_{i}^{cont}\sim\mathcal{N}(\mathbf{50},10^2\Sigma)$ & MM & 1.39 & 1.40 & 1.41 & 1.44 & 1.50 \\
   &  &  &  &  &  &  \\
   & shooting S + BI & 1.70 & 1.66 & 1.65 & 1.67 & 1.76 \\
   & shooting S + skH & 2.00 & 1.91 & 1.88 & 1.79 & 1.86 \\
   \hline
\end{tabular}
}
\end{table}

For rowwise contamination the situation is different (see Table \ref{tab:res_row_MSE_corrpred}). Here, as known, MM and S-estimation give excellent results. The shooting S-estimators give only slightly higher values of MSE compared to the ordinary S-estimator, indicating that the shooting S-estimators can cope with rowwise contamination as well. Nevertheless, as the shooting S-estimator has been developed for cellwise contamination, we do not advise its usage if there is only rowwise contamination present.

\begin{table}[t]
\centering
\caption{\small{$n\cdot MSE$ of different estimators for vertical outliers with $n=100$, $p=15$ and predictors with correlation matrix $\Sigma$}}
\label{tab:res_vertical}
{\footnotesize
\begin{tabular}{|r|rrrrr|}
  \hline
 & $\epsilon = 0$ & $\epsilon = 0.01$ & $\epsilon = 0.02$ & $\epsilon = 0.05$ & $\epsilon = 0.1$ \\
  \hline
LS & 1.28 & 51.40 & 97.69 & 234.23 & 438.65 \\
  S & 1.53 & 1.51 & 1.50 & 1.48 & 1.48 \\
  MM & 1.39 & 1.40 & 1.41 & 1.44 & 1.50 \\
   &  &  &  &  &  \\
  shooting S +  BI & 1.70 & 1.73 & 1.77 & 1.88 & 2.13 \\
  shooting S + skH & 2.00 & 2.03 & 2.01 & 2.10 & 2.14 \\
   \hline
\end{tabular}
}
\end{table}

The shooting S-estimator can also cope with vertical outliers (see Table \ref{tab:res_vertical}). It gives good results for all levels of contamination used here, although its MSE is slightly higher than for the S- and MM-estimators. The reason for the good performance of the shooting S-estimator is that the contamination in the response is present in the computation of each single coefficient $\hat{\beta}_j$. Robustness of the `regression step' leads to small weights $w_{ij}$ for all $j$.
We may conclude that the shooting S-estimator is the only considered regression estimator that can deal with cellwise contamination above $2\%$ in our simulation setting. The estimator also gives good results in presence of vertical outliers. If there are no outliers, there is a slight loss in efficiency compared to the other robust estimators.  In a rowwise contamination setting, we advise the use of the usual  S- and MM-estimator.


\section{Real Data}
\label{sec:realdata}

We evaluate the performance of the shooting S-estimator on real data sets and compare it to the LS, S- and MM-estimators. For all estimators, the tuning parameters are chosen as in Section~\ref{sec:sim}. We choose the three data sets \texttt{Cars93}, \texttt{Auto} and \texttt{Boston}. When applying the shooting S-estimator, we declare a component of an observation, hence a cell in the data matrix, as an outlier if it gets a robustness weight below $0.5$. If all components of an observation are flagged as outliers, we say that the whole observation is outlying.

The \texttt{Cars93} data, a selection of 1993 model cars, are included in the \textsf{R} package \texttt{MASS}. Omitting not fully observed data points, we are left with $n=82$ observations. We fit the following model with $p=14$ predictor variables of the \texttt{Cars93} data (for the definition of the variables, see Table \ref{tab:Cars93_descr} in the appendix)
\begin{align*}
PRICE = & \beta_0 + \beta_1 MPG.C + \beta_2 MPG.H + \beta_3 ENG.SIZE + \beta_4 HP \\
& + \beta_{5} RPM + \beta_{6} REV.MILE + \beta_{7} FUEL.TANK  + \beta_{8} LENGTH \\
& + \beta_{9} WHEELBASE + \beta_{10} WIDTH + \beta_{11} TURN  + \beta_{12} REAR.SEAT \\
& + \beta_{13} LUGGAGE + \beta_{14} WEIGHT  + error.
\end{align*}
 The shooting S-estimator using a biweight loss downweights {seven} observations as a whole and detects outlying cells for another {19} observations. In contrast, the MM-estimator  downweights the observations corresponding to these outlying cells as a whole, thereby loosing information. This information loss is especially visible when looking, for example, at observation 46, which receives the weight 0 by the MM-estimator, while the shooting S-estimator with biweight loss assigns a weight of about 1 to all components except the first component, which receives weight 0.

The \texttt{Auto} data set is included in Stata and can be downloaded from \url{http://www.stata-press.com/data/r13/auto.dta} \citep[see][]{StataManual}. It consists of $n=74$ fully observed sales of vintage 1978 automobiles in the United States (see Table \ref{tab:auto_descr} in the appendix). We fit the following model with $p=8$ predictor variables
\begin{align*}
PRICE = &\beta_0 + \beta_1 MPG + \beta_2 HEADROOM + \beta_3 TRUNK  + \beta_4 WEIGHT \\
& + \beta_5 LENGTH + \beta_6 TURN + \beta_7 DISPLACE + \beta_8 GEAR + error.
\end{align*}
The shooting S-estimator with biweight loss downweights five observation as a whole and flags cells of another 17 observations as outliers. For instance, observations 12 (`Chevrolet Cavalier') and 13 (`Chevrolet Corsica') receive a weight of zero by MM {and the ordinary S}, while the shooting S-estimator using a biweight loss finds out that only component 2,  the headroom, is outlying.  Again, we  conclude that the shooting S-estimator uses more information from the data than the MM-estimator or the ordinary S-estimator.

The third data set, the \texttt{Boston} housing data, originates from \cite{Harrison} and has been extensively analyzed in the robust statistics literature. The data are available in the \textsf{R} package \texttt{mlbench} and contain various characteristics of houses, demographics, air pollution and geographical details on $n=506$ census tracts in and nearby Boston. Table \ref{tab:boston_descr} in the appendix gives an overview of the definition of the variables ($p=9$) in the model
\begin{align*}
\log(MEDV) = &\beta_0 + \beta_1 CRIM + \beta_2 NOX^2 + \beta_3 RM^2 + \beta_4 AGE + \beta_5 \log(DIS) \\
& + \beta_6 TAX + \beta_7 PTRATIO + \beta_8 B + \beta_9 \log(LSTAT) + error.
\end{align*}
\cite{Belsley} discovered outlying behavior of census tracts lying in central area of Boston, concentrated in three neighborhoods. Applying the shooting S-estimator using a biweight loss to the full data set, we get similar results. The shooting S-estimator declares the observations from the neighborhoods Back Bay ($365-370$), Beacon Hill (371- 373) and South Boston (394-406) as cellwise contaminated, with mainly the components corresponding to  the variables \texttt{RM} and \texttt{AGE}  indicated as outlying. The MM-estimator and the ordinary S-estimator downweight as a whole the observations of the neighborhoods Back Bay and Beacon Hill and half of the observations of South Boston, resulting in a loss of information.
\vspace{\baselineskip}

 For each of the three data sets, we randomly choose $4/5$th of the observations and compute all estimates on this training data set. This we repeat $R=500$ times and we compare the estimates on the training data sets $\mathbf{\hat{\beta}}^{(r)}$, for $r=1,\ldots,R$, to the one computed on the full data set $\mathbf{\hat{\beta}}^{full}$. Adjusting for the different scales of the explanatory variables, we get what we call  the Average Norm Distance (AND)
\begin{align}
\text{AND}(\mathbf{\hat{\beta}}) =\frac{1}{R} \sum_{r=1}^R \sqrt{\frac{1}{p}\sum_{j=1}^p (\hat{\beta}^{(r)}_j - \hat{\beta}_j^{full})^2  \frac{\MAD(x_{1j},\ldots,x_{nj})^2}{\MAD(y_1,\ldots, y_n)^2}}.
\label{eq:AND}
\end{align}
A low value of AND is desired. Table \ref{tab:AND} shows the results for  all considered estimators on the three data sets. As pointed out by a referee, the AND criterion is a version of the Jackknife estimator of the variance, and it reflects the efficiency. Therefore, the low value of AND for the LS estimator is no surprise. The AND for the S- and MM-estimator are close to those of LS, and sometimes even slightly better. The shooting S-estimators have somehow larger values of the AND, but the loss in efficiency remains limited.

\begin{table}
\centering
\caption{\small{Average norm distance ($AND$) for five estimators computed on three data sets and their contaminated versions}} 
\label{tab:AND}
{\footnotesize
\begin{tabular}{|c|rrr|rrr|}
\hline 
&\multicolumn{3}{|c|}{observed data} & \multicolumn{3}{|c|}{contaminated data}\\
  \hline
 & Auto & Cars93 & Boston & Auto & Cars93 & Boston \\ 
  \hline
LS & 0.388 & 0.141 & 0.024 & 1.320 & 0.325 & 0.273 \\ 
  S & 0.459 & 0.172 & 0.021 & 0.697 & 0.240 & 0.223 \\ 
  MM & 0.282 & 0.213 & 0.022 & 0.346 & 0.243 & 0.179 \\ 
  shooting S + BI & 0.607 & 0.217 & 0.033 & 0.251 & 0.228 & 0.152 \\ 
  shooting S + skH & 0.574 & 0.186 & 0.039 & 0.658 & 0.169 & 0.138 \\ 
   \hline
\end{tabular}
}
\end{table}

To investigate the robustness of the estimators, we randomly choose $5$\% of the cells of the data set and replace them with $x_{ij}^{cont}\sim\mathcal{N}(\hat{\mu}_j+10\hat{\sigma}_j, \hat{\sigma}_j^2)$ where $\hat{\mu}_j$ and $\hat{\sigma}_j$ denote the median and MAD of the $j$th column of the design matrix, respectively. This we repeat $R=500$ times and we compute the average norm difference as in (\ref{eq:AND}), where $\mathbf{\hat{\beta}}^{(r)}$  are the estimates from the contaminated data  and $\mathbf{\hat{\beta}}^{full}$ is the estimate on the original data.  Table~\ref{tab:AND} gives the results. Now the AND measures the robustness of the estimators, and the  LS estimator clearly gives the worst results. The shooting S-estimators, and in particular when using  a biweight loss, give  the best results. They deal better with
 cellwise contamination than the ordinary S- and MM-estimator.

We did not use a  prediction error criterion to assess the performance of the different estimators. If the level of cellwise contamination is moderate to high, we expect that most observations contain  contaminated cells.  When forecasting, outlying components of an observation get full influence (which is not the case in  estimation).  Assessing the  prediction error by cross-validation is then not reliable anymore, since the validation set contains too many  observations with contaminated components. Using a robust cross-validation criterion, as a trimmed mean squared prediction error does not solve this problem, as far too many observations used for validation may have outlying components.   Prediction for cell-wise contaminated observations is left as a topic for future research.


\section{Conclusion}
\label{sec:conc}

In this paper, we introduce a regression estimator applicable for cellwise contamination. It combines the ideas of ordinary regression S-estimation with the coordinate descent algorithm. Thereby the shooting S-estimator is able to use different weights for different components of an observation. In our simulations, it can deal with cellwise contamination up to 10\%.

Furthermore, the shooting S-estimator can also be used as a diagnostic tool. After computation of the shooting S-estimate, the entries of the weight matrix $w_{ij}$ help to distinguish between clean data and outliers, and even between cellwise and rowwise contamination. While high weights indicate a clean cell, low weights indicate contamination. If all components of the same observation get low weights, this means that all components are contaminated or that it is a vertical outlier.

The efficiency of the shooting S-estimator  can be improved by using a \textit{shooting MM-estimator} instead. To obtain a shooting MM-estimator, the simple S-estimation step inside the algorithm  needs to be replaced with a simple MM-estimation. In order to explore this idea, the simulations of Section~\ref{sec:sim} were repeated for a shooting MM-estimator, using simple MM-estimation with 20\% breakdown point and 95\% efficiency at the normal distribution. Preliminary results indicate that (i) the  shooting  MM-estimator gave generally lower values for mean squared error in the simulations of Section~\ref{sec:sim} than the shooting S-estimator; (ii) especially for clean data and small amounts of contamination, the improvement of the shooting MM-estimator over the shooting S-estimator was clearly visible; (iii) the shooting MM-estimator outperformed the ordinary MM-estimator in any setting where cellwise contamination was present. However, further development of the shooting MM-estimator is necessary and is left for future research.

Another idea worth considering is the application of an imputation method after performing shooting S-regression. Cells that are flagged as outliers can be set as missing. On the data set containing missing values, regression can be performed \citep[see][]{Little}.

Admittedly, our shooting S-estimator has problems with cellwise good leverage points, which are observations with large values in some single cells that do follow the regression model. The shooting S-estimator tends to flag the contaminated cells of the good leverage points as outliers when computing the starting values of the algorithm. However, if the data contain rowwise good leverage points, thus large values for all cells of {observations that do} follow the model, the shooting S-estimator behaves comparable to the other estimators (LS, S, MM) in our experiments.

As the shooting S-algorithm deals with each variable separately, it can also be applied to data sets with a small sample size and even if $n<p$. A suitable $\rho$-function for this setting may be the linear quadratic quadratic (lqq) function of \cite{Koller}, as it has been shown to have high efficiency also for small sample sizes. When using the lqq-function in the simulation setups in Section \ref{sec:sim}, the results are comparable to the ones with the other $\rho$-functions used there.

Finally, the shooting S-estimator can be extended to a \textit{penalized shooting S-estimator}. To the simple S-estimation in every variable, a penalty term $J(\beta)$ can be added. Possible choices for the penalty term are $J(\beta) = |\beta|$ or $J(\beta) = \beta^2$. The penalized version of the shooting S-estimator could be very useful in high-dimensional settings.
\bigskip

\textbf{Acknowledgements} \small{We  gratefully acknowledge support from the GOA/12/014 project of the Research Fund KU Leuven.
We thank the referees for their constructive comments, and in particular the third anonymous referee who corrected some flaws in the first version of the paper and who made many suggestions for improving the write up of the paper.
}

\newpage
\appendix
\section{APPENDIX - Description of Variables for Real Data Examples}

\begin{table}[h]
\scriptsize
\centering
\caption{Variables of the \texttt{Cars93} data}
\label{tab:Cars93_descr}
\begin{tabular}{|l|l|}
\hline
Name & Description\\
\hline
$PRICE$ & Midrange Price (in \$1,000)\\
$MPG.C$ & City MPG (miles per US gallon by EPA rating)\\
$MPG.H$ & Highway MPG (miles per US gallon by EPA rating)\\
$ENG.SIZE$ & Engine displacement size in liters\\
$HP$ & Maximum horsepower\\
$RPM$ & Revolutions per minute at which maximum horsepower is achieved\\
$REV.MILE$ & Number of revolutions of the engine needed for car to travel one\\
& mile in its highest gear\\
$FUEL.TANK$ & Capacity of the fuel tank in US gallons\\
$LENGTH$ & Length of the car in inches\\
$WHEELBASE$ & Size of the wheelbase in inches\\
$WIDTH$ & Width of the car in inches\\
$TURN$ & U-turn space in feet\\
$REAR.SEAT$ & Rear seat room in inches\\
$LUGGAGE$ & Luggage capacity in cubic feet\\
$WEIGHT$ & Weight of the car in pounds\\
\hline
\end{tabular}
\end{table}

\begin{table}[h]
\centering
\scriptsize
\caption{Variables of the \texttt{Auto} data}
\label{tab:auto_descr}
\begin{tabular}{|l|l|}
\hline
Name & Description\\
\hline
$PRICE$ & Price in US-dollars\\
$MPG$ & Milage\\
$HEADROOM$ & Head room in inches\\
$TRUNK$ & Trunk space in cubic feet\\
$WEIGHT$ & Weight of the car in pounds\\
$LENGTH$ & Length of the car in inches\\
$TURN$ & U-turn space in feet\\
$DISPLACE$ & Displacement in cubic inches\\
$GEAR$ & Gear ratio\\
\hline
\end{tabular}
\end{table}

\begin{table}[h]
\centering
\caption{Variables of the \texttt{Boston} data}
\label{tab:boston_descr}
\scriptsize
\begin{tabular}{|l|l|}
\hline
Name & Description\\
\hline
$MEDV$ & Median value of owner-occupied homes in USD 1000's\\
$CRIM$ & Per capita crime rate by town\\
$NOX$ & Nitric oxides concentration in parts per 10 million\\
$RM$ & Average number of rooms per dwelling\\
$AGE$ & Proportion of owner-occupied units built prior to 1940\\
$DIS$ & Weighted distance to five Boston employment centres\\
$TAX$ & Full-value property-tax rate per USD 10,000\\
$PTRATIO$ & Pupil-Teacher ratio by town\\
$B$ & Proportion of black population\\
$LSTAT$ & Percentage of lower status population\\
\hline
\end{tabular}
\end{table}

\clearpage
\bibliographystyle{plainnat}
\bibliography{cellwiseRevision3}

\end{document}